# Circulant Arrays on Cyclic Subgroups of Finite Fields: Rank Analysis and Construction of Quasi-Cyclic LDPC Codes

Li Zhang[1], Shu Lin[1], Khaled Abdel-Ghaffar[1], Zhi Ding[1], and Bo Zhou[2]

[1]Department of Electrical and Computer Engineering, University of California

Davis, CA 95616

[2]Qualcomm Inc., San Diego, CA 92121

Email: liszhang@ece.ucdavis.edu, shulin@ece.ucdavis.edu,

ghaffar@ece.ucdavis.edu, zding@ece.ucdavis.edu and bzhou@qualcomm.com

### Abstract

This paper consists of three parts. The first part presents a large class of new binary quasi-cyclic (QC)-LDPC codes with girth of at least 6 whose parity-check matrices are constructed based on cyclic subgroups of finite fields. Experimental results show that the codes constructed perform well over the binary-input AWGN channel with iterative decoding using the sum-product algorithm (SPA). The second part analyzes the ranks of the parity-check matrices of codes constructed based on finite fields with characteristic of 2 and gives combinatorial expressions for these ranks. The third part identifies a subclass of constructed QC-LDPC codes that have large minimum distances. Decoding of codes in this subclass with the SPA converges very fast.

## I. INTRODUCTION

The rapid dominance of LDPC codes [1] in applications requiring error control coding is due to their capacity-approaching performance. LDPC codes were first discovered by Gallager

This research was supported by NSF under the Grant CCF-0727478 and NASA under the Grant NNX09AI21G and gift grants from Northrop Grumman Space Technology, Intel and Denali Software Inc.

in 1962 [1] and then rediscovered in late 1990's [2], [3]. Ever since their rediscovery, a great deal of research effort has been expended in design, construction, structural analysis, encoding, decoding, performance analysis, generalizations and applications of LDPC codes. Many LDPC codes have been adopted as the standard codes for various next generations of communication systems.

A regular binary LDPC code [1] is given by the null space of a sparse parity-check matrix $\mathbf{H}$ over GF(2) with constant column weight $\gamma$ and constant row weight $\rho$. Such an LDPC code is said to be $(\gamma, \rho)$-regular. If the columns and/or rows of $\mathbf{H}$ have multiple weights, then the null space of $\mathbf{H}$ gives an irregular LDPC code. If $\mathbf{H}$ is an array of sparse circulants of the same size over GF(2), then the null space of $\mathbf{H}$ gives a binary quasi-cyclic (QC)-LDPC codes.

In almost all of the proposed constructions of LDPC codes, the following constraint on the rows and columns of the parity-check matrix $\mathbf{H}$ is imposed: *no two rows (or two columns) can have more than one place where they both have 1-components*. This constraint on the rows and columns of $\mathbf{H}$ is referred to as the *row-column (RC)-constraint*. The RC-constraint ensures that the Tanner graph [4] of the LDPC code given by the null space of $\mathbf{H}$ has a girth of at least 6 [5], [6]. It also ensures that the minimum distance of a $(\gamma, \rho)$-regular LDPC code is at least $\gamma + 1$. This distance bound is tight for regular LDPC codes whose parity-check matrices have large column weights, such as finite geometry cyclic LDPC codes [5] and finite field QC-LDPC codes constructed in [7]-[9] and this paper. A parity-check matrix that satisfies the RC-constraint is called an *RC-constrained parity-check matrix*.

This paper is concerned with construction of algebraic QC-LDPC codes. QC-LDPC codes can be efficiently encoded using simple shift-registers [10]. Furthermore, for hardware decoder implementation, their QC-structure simplifies wire routing [11] and allows partially parallel decoding [12] which offers a trade-off between decoding complexity and decoding speed. Well designed algebraic QC-LDPC codes can perform close to the Shannon limit and just as well as or even better than their corresponding random or pseudo-random QC-LDPC codes constructed using computer-based methods over the binary-input AWGN and binary erasure channel (BEC), as demonstrated in [7]-[9]. In [7], a general and three specific methods for constructing algebraic QC-LDPC codes based on finite fields were presented.

In this paper, we present a new class of RC-constrained matrices constructed based on cyclic subgroups of finite fields. Based on this class of RC-constrained matrices, a large class of RC-constrained QC-LDPC codes is constructed using the general method presented in [7]. Also in this paper, we analyze the ranks of the parity-check matrices of the codes in several subclasses of the new codes. Furthermore, we identify a subclass of RC-constrained QC-LDPC codes that have large minimum distances.

The rest of this paper is organized as follows. In Section II, we give a brief review of the general method for constructing algebraic QC-LDPC codes presented in [7]. In Section III, we first present a large class of new RC-constrained matrices and then give a new class of algebraic RC-constrained QC-LDPC codes. The construction of this new class of RC-constrained QC-LDPC codes is based on *cyclic subgroups* of finite fields. We show that this new class of QC-LDPC codes contains the major class of QC-LDPC codes constructed in [7] (method-1) as a subclass. We also show that this new class of codes is a large expansion of the QC-LDPC codes constructed by the third method given in [8]. In Section IV, we analyze the ranks of the parity-check matrices of codes in several subclasses of the new codes. In Section V, we characterize a special subclass of new RC-constrained QC-LDPC codes that have large minimum distances and are effective for a reliability-based iterative decoding algorithm for a trade-off between error performance and decoding complexity. Section VI concludes the paper with some remarks.

The construction of RC-constrained QC-LDPC codes based on cyclic subgroups of finite fields presented in this paper is a counter part of the construction of RC-constrained QC-LDPC codes based on *additive subgroups* of finite fields presented in [9]. In [9], no rank analysis of the parity-check matrices of codes is provided.

## II. A GENERAL ALGEBRAIC CONSTRUCTION OF QC-LDPC CODES

Consider the Galois field GF($q$) where $q$ is a power of a prime. Let $\alpha$ be a primitive element of GF($q$). Then, the powers of $\alpha$, $\alpha^{-\infty} \triangleq 0, \alpha^0 = 1, \alpha, \alpha^2, \ldots, \alpha^{q-2}$, give all the $q$ elements of GF($q$) and $\alpha^{q-1} = 1$. The $q-1$ nonzero elements of GF($q$) form the multiplicative group of GF($q$) under the multiplicative operation defined on GF($q$).

Let $\mathbf{P}$ be a $(q-1) \times (q-1)$ circulant permutation matrix (CPM) whose top row is given by the $(q-1)$-tuple $(0\,1\,0\ldots0)$ over GF(2) where the components are labeled from 0 to $q-2$ and the single 1-component is located at the position labeled by "1". Then $\mathbf{P}$ consists of the $(q-1)$-tuple $(0\,1\,0\ldots0)$ and its $q-2$ right cyclic-shifts as rows. For $1 \le i < q$, let $\mathbf{P}^i$ be the product of $\mathbf{P}$ with itself $i$ times, called the $i$th power of $\mathbf{P}$. Then, $\mathbf{P}^i$ is also a $(q-1) \times (q-1)$ CPM whose top row has a single 1-component at the $i$th position. For $i = q - 1$, $\mathbf{P}^{q-1} = \mathbf{I}_{q-1}$, the $(q-1) \times (q-1)$ identity matrix. Let $\mathbf{P}^0 \triangleq \mathbf{P}^{q-1} = \mathbf{I}_{q-1}$. Then the set $\mathscr{P} = \{\mathbf{P}^0, \mathbf{P}, \mathbf{P}^2, \ldots, \mathbf{P}^{q-2}\}$ of CPMs forms a cyclic group of order $q - 1$ under matrix multiplication over GF(2) with $\mathbf{P}^{q-1-i}$ as the multiplicative inverse of $\mathbf{P}^i$ and $\mathbf{P}^0$ as the identity element.

For the nonzero element $\alpha^i$ in GF($q$) with $0 \leq i < q-1$, we represent it by the $(q-1) \times (q-1)$ CPM $\mathbf{P}^i$ in $\mathscr{P}$. This matrix representation is referred to as the $(q-1)$-*fold binary matrix dispersion* (or simply binary matrix dispersion) of $\alpha^i$. It is clear that the binary matrix dispersions of two different nonzero elements in GF($q$) are different. Since there are exactly $q - 1$ different $(q-1) \times (q-1)$ CPMs in $\mathscr{P}$, there is a one-to-one correspondence between a nonzero element of GF($q$) and a $(q-1) \times (q-1)$ CPM in $\mathscr{P}$. Therefore, each nonzero element of GF($q$) is uniquely represented by a $(q-1) \times (q-1)$ CPM in $\mathscr{P}$. For a nonzero element $\delta$ in GF($q$), we use the notation $\mathbf{B}(\delta)$ to denote its binary matrix dispersion. If $\delta = \alpha^i$, then $\mathbf{B}(\delta) = \mathbf{P}^i$. For the 0-element of GF($q$), its binary matrix dispersion is defined as the $(q-1) \times (q-1)$ zero matrix (ZM), denoted by $\mathbf{P}^{-\infty}$.

Consider a $k \times n$ matrix over GF($q$),

$$\mathbf{W} = \begin{bmatrix} \mathbf{w}_0 \\ \mathbf{w}_1 \\ \vdots \\ \mathbf{w}_{k-1} \end{bmatrix} = \begin{bmatrix} w_{0,0} & w_{0,1} & \cdots & w_{0,n-1} \\ w_{1,0} & w_{1,1} & \cdots & w_{1,n-1} \\ \vdots & \vdots & \ddots & \vdots \\ w_{k-1,0} & w_{k-1,1} & \cdots & w_{k-1,n-1} \end{bmatrix}, \tag{1}$$

whose rows satisfy the following constraint: for $0 \leq i, j < k, i \neq j$ and $0 \leq c, l < q - 1$, the Hamming distance between the two $q$-ary $n$-tuples, $\alpha^c \mathbf{w}_i$ and $\alpha^l \mathbf{w}_j$, is at least $n - 1$, (i.e., $\alpha^c \mathbf{w}_i$ and $\alpha^l \mathbf{w}_j$ differ in at least $n - 1$ places). The above constraint on the rows of matrix $\mathbf{W}$ is called the *row-distance (RD)-constraint* and $\mathbf{W}$ is called an RD-constrained matrix.

For $0 \leq i < k$ and $0 \leq j < n$, dispersing each nonzero entry $w_{i,j}$ of $\mathbf{W}$ into a $(q-1) \times (q-1)$ CPM $\mathbf{B}_{i,j} \triangleq \mathbf{B}(w_{i,j})$ over GF(2) and each 0-entry into a $(q-1) \times (q-1)$ ZM, we obtain the following $k \times n$ array (or block) of CPMs and/or ZMs over GF(2) of size $(q-1) \times (q-1)$:

$$\mathbf{H} = [\mathbf{B}_{i,j}]_{0 \leq i < k, 0 \leq j < n}. \tag{2}$$

$\mathbf{H}$ is called the binary $(q-1)$-fold *array dispersion* of $\mathbf{W}$ (or simply binary array dispersion of $\mathbf{W}$) and it is a $k(q-1) \times n(q-1)$ matrix over GF(2). Based on the RD-constraint on the rows of $\mathbf{W}$ and the binary CPM matrix dispersions of the entries of $\mathbf{W}$, it was proved in [7],[8] that $\mathbf{H}$, as a $k(q-1) \times n(q-1)$ matrix over GF(2), satisfies the RC-constraint.

The null space of $\mathbf{H}$ gives a binary RC-constrained QC-LDPC code $\mathcal{C}_{qc}$ of length $n(q-1)$ whose Tanner graph has a girth of at least 6. The subscript "$qc$" of $\mathcal{C}_{qc}$ stands for "quasi-cyclic". If $\mathbf{H}$ has constant column and row weights, then $\mathcal{C}_{qc}$ is a regular QC-LDPC code; otherwise, it is an irregular QC-LDPC code. Since $\mathbf{H}$ is an array of CPMs and ZMs, it is referred to as a *circulant-based parity-check matrix* (CPCM). Any RC-constrained CPCM gives a QC-LDPC code with girth at least 6. In [7]-[9] several classes of RD-constrained matrices over finite fields were given. By array dispersions of these classes of RD-constrained matrices, several classes of RC-constrained QC-LDPC codes were constructed. The codes given in the examples of [7]-[9] decoded with iterative decoding using the sum-product algorithm (SPA) displayed excellent performance in terms of error-rate, error-floor and rate of decoding convergence.

## III. A Class of RC-Constrained QC-LDPC Codes on Finite Fields

In this section, we first present a large and very flexible class of RD-constrained matrices constructed based on cyclic subgroups of finite fields. Then, based on this class of RD-constrained matrices, we construct a class of RC-constrained QC-LDPC codes.

### A. A Class of RD-Constrained Matrices

Let $\alpha$ be a primitive element of GF($q$). Suppose that $q-1$ can be factored as a product of two integers, $c$ and $n$, that are relatively prime. Then $q-1 = cn$. Let $\beta = \alpha^c$ and $\delta = \alpha^n$. Then the orders of $\beta$ and $\delta$ are $n$ and $c$, respectively. The set $\mathscr{G}_1 = \{\beta^0 = 1, \beta, \ldots, \beta^{n-1}\}$ and the set $\mathscr{G}_2 = \{\delta^0 = 1, \delta, \ldots, \delta^{c-1}\}$ form two cyclic subgroups of the multiplicative group of GF($q$). Since $c$ and $n$ are relatively prime, $\mathscr{G}_1$ and $\mathscr{G}_2$ can only have the unit element "1" in common.

For $0 \le i, j < c$, form the following $n \times n$ matrix over GF($q$) using a single element from $\mathscr{G}_2$ and all the elements in $\mathscr{G}_1$:

$$\mathbf{W}_{i,j} = \begin{bmatrix} \delta^{j-i}\beta^0 - \beta^0 & \delta^{j-i}\beta^0 - \beta^1 & \cdots & \delta^{j-i}\beta^0 - \beta^{n-1} \\ \delta^{j-i}\beta^1 - \beta^0 & \delta^{j-i}\beta^1 - \beta^1 & \cdots & \delta^{j-i}\beta^1 - \beta^{n-1} \\ \vdots & \vdots & \ddots & \vdots \\ \delta^{j-i}\beta^{n-1} - \beta^0 & \delta^{j-i}\beta^{n-1} - \beta^1 & \cdots & \delta^{j-i}\beta^{n-1} - \beta^{n-1} \end{bmatrix}. \tag{3}$$

From the structure of $\mathbf{W}_{i,j}$ displayed by (3), we can readily see or prove that $\mathbf{W}_{i,j}$ has the following structural properties: 1) each row is the right cyclic-shift of the row above it multiplied by $\beta$ and the first row is the right cyclic-shift of the last row multiplied by $\beta$; 2) each column is the downward cyclic-shift of the column on its left multiplied by $\beta$ and the first column is the downward cyclic-shift of the last column multiplied $\beta$; 3) all the entries in a row (or a column) are distinct elements of GF($q$); 4) any two rows (or columns) differ in every position; 5) for $i \neq j$, all the entries in $\mathbf{W}_{i,j}$ are nonzero elements of GF($q$); and 6) for $i = j$, the entries on the main diagonal of $\mathbf{W}_{i,i}$ are zeros and all the other entries are nonzero.

*Theorem 1:* For $0 \leq i, j < c$, the $n \times n$ matrix given by (3) $\mathbf{W}_{i,j}$ satisfies the RD-constraint.

*Proof:* Let $\mathbf{w}_k$ and $\mathbf{w}_l$ be two different rows in $\mathbf{W}_{i,j}$. Then, $k \neq l$. For any two integers $e$ and $f$ with $0 \leq e, f < q - 1$, consider the two $n$-tuples over GF($q$), $\alpha^e \mathbf{w}_k$ and $\alpha^f \mathbf{w}_l$. It follows from the structural properties 4 to 6 of $\mathbf{W}_{i,j}$ that $\alpha^e \mathbf{w}_i$ and $\alpha^f \mathbf{w}_j$ cannot have any position where they both have 0-components. Next, we prove that $\alpha^e \mathbf{w}_k$ and $\alpha^f \mathbf{w}_l$ cannot have more than one position where they have identical nonzero components. Suppose that $\alpha^e \mathbf{w}_k$ and $\alpha^f \mathbf{w}_l$ have identical nonzero components at two different positions $s$ and $t$ ($s \neq t$) . Then, we have the following equalities: $\alpha^e (\delta^{j-i} \beta^k - \beta^s) = \alpha^f (\delta^{j-i} \beta^l - \beta^s)$ and $\alpha^e (\delta^{j-i} \beta^k - \beta^t) = \alpha^f (\delta^{j-i} \beta^l - \beta^t)$. From these two equalities, we obtain the equality $(\beta^t - \beta^s)(\beta^l - \beta^k) = 0$. This equality holds if and only if either $k = l$ or $s = t$ which contradicts the facts that neither $k \neq l$ nor $s \neq t$. Therefore, $\alpha^e \mathbf{w}_k$ and $\alpha^f \mathbf{w}_l$ can not have more than one position where they have identical nonzero components. It follows from the above proven facts that $\mathbf{W}_{i,j}$ satisfies the RD-constraint. ∎

It follows from Theorem 1 that $\Omega = \{\mathbf{W}_{i,j} : 0 \leq i, j < c\}$ gives a set of $c^2$ RD-constrained matrices over GF($q$). Each matrix in $\Omega$ can be used as a base matrix for array dispersion to construct QC-LDPC codes. In the following, we show that the RD-constrained matrices in $\Omega$ can be used to form a much larger RD-constrained matrix for array dispersion to construct QC-LDPC codes.

Form the following $c \times c$ array with $\mathbf{W}_{i,j}$, $0 \leq i, j < c$, as sub-matrices:

$$\mathbf{W} = \begin{bmatrix} \mathbf{W}_{0,0} & \mathbf{W}_{0,1} & \cdots & \mathbf{W}_{0,c-1} \\ \mathbf{W}_{1,0} & \mathbf{W}_{1,1} & \cdots & \mathbf{W}_{1,1} \\ \vdots & \vdots & \ddots & \vdots \\ \mathbf{W}_{c-1,0} & \mathbf{W}_{c-1,1} & \cdots & \mathbf{W}_{c-1,c-1} \end{bmatrix} . \tag{4}$$

$\mathbf{W}$ is a $c \times c$ array of $n \times n$ sub-matrices. Since $cn = q - 1$, $\mathbf{W}$ is a $(q-1) \times (q-1)$ matrix over GF($q$). For $0 \le i, j < c$, it follows from the composition of the entries of $\mathbf{W}_{i,j}$s displayed in (3), we readily see that $\mathbf{W}_{i+1,j+1} = \mathbf{W}_{i,j}$ with $i+1$ and $j+1$ reduced by modulo-$c$. Then, every row of submatrices of $\mathbf{W}$ is a right cyclic-shift of the row above it and the first row is the right cyclic-shift of the last row.

For $0 \le i < c, 0 \le k < n$, every integer in $\{0, 1, 2, \ldots, cn - 1 = q - 2\}$ can be expressed as $in + k$. Let $\mathbf{w}_{in+k}$ denote the $(in + k)$th row of $\mathbf{W}$, as a $(q-1) \times (q-1)$ matrix. Then

$$\mathbf{w}_{in+k} = (\mathbf{w}_{i,0,k}, \mathbf{w}_{i,1,k}, \ldots, \mathbf{w}_{i,c-1,k}), \tag{5}$$

which consists of $c$ sections, $n$ components each. For $0 \le j < c$, the $j$th section $\mathbf{w}_{i,j,k} = (\delta^{j-i}\beta^k - \beta^0, \delta^{j-i}\beta^k - \beta^1, \ldots, \delta^{j-i}\beta^k - \beta^{n-1})$ of $\mathbf{w}_{in+k}$ is simply the $k$th row of the submatrix $\mathbf{W}_{i,j}$ of $\mathbf{W}$. From (5) and properties 5 and 6 of each submatrix $\mathbf{W}_{i,j}$ of $\mathbf{W}$, we readily see that $\mathbf{w}_{in+k}$ contains one and only one 0-component at the $(in + k)$th position (or the $k$th position of $i$th section $\mathbf{w}_{i,i,k}$). Therefore, $\mathbf{W}$ contains $q - 1$ 0-entries that lie on the main diagonal of $\mathbf{W}$, as a $(q-1) \times (q-1)$ matrix over GF($q$). It follows from property 4 of each submatrix $\mathbf{W}_{i,j}$ that any two rows of $\mathbf{W}$ differ in every position.

*Theorem 2:* The $(q-1) \times (q-1)$ matrix $\mathbf{W}$ given by (4) satisfies the RD-constraint.

*Proof:* Let $0 \le i_1, i_2 < c, 0 \le k_1, k_2 < n$, and $i_1 n + k_1 \ne i_2 n + k_2$. In this case, either $i_1 \ne i_2$ or $k_1 \ne k_2$. Then $\mathbf{w}_{i_1 n + k_1}$ and $\mathbf{w}_{i_2 n + k_2}$ are two different rows of $\mathbf{W}$. It follows from structural property 4 of the RD-constrained submatrices $\mathbf{W}_{i,j}$s that $\mathbf{w}_{i_1 n + k_1}$ and $\mathbf{w}_{j_2 n + k_2}$ differ in every position and cannot have any position where they both have 0-components. For $0 \le e, f < q - 1$, consider the two $q$-ary $(q-1)$-tuples, $\alpha^e \mathbf{w}_{i_1 n + k_1}$ and $\alpha^f \mathbf{w}_{i_2 n + k_2}$. Suppose there are two different positions, $j_1 n + s$ and $j_2 n + t$ with $0 \le j_1, j_2 < c, 0 \le s, t < n$ (i.e., $j_1 n + s \ne j_2 n + t$), where $\alpha^e \mathbf{w}_{i_1 n + k_1}$ and $\alpha^f \mathbf{w}_{i_2 n + k_2}$ have identical nonzero components. Based on this hypothesis, we have following two equalities: $\alpha^e(\delta^{j_1 - i_1}\beta^{k_1} - \beta^s) = \alpha^f(\delta^{j_1 - i_2}\beta^{k_2} - \beta^s)$, and $\alpha^e(\delta^{j_2 - i_1}\beta^{k_1} - \beta^t) = \alpha^f(\delta^{j_2 - i_2}\beta^{k_2} - \beta^t)$. From these two equalities with some algebraic manipulations, we obtain the following equality: $\delta^{j_2 - j_1} = \beta^{t-s}$. Since $\beta$ and $\delta$ are elements in the cyclic subgroups $\mathscr{G}_1$ and $\mathscr{G}_2$, respectively, and $\mathscr{G}_1 \bigcap \mathscr{G}_2 = \{1\}$, the equality $\delta^{j_2 - j_1} = \beta^{t-s}$ holds if and only if $j_2 = j_1$ and $t = s$ simultaneously. These two equalities imply that $j_1 n + s = j_2 n + t$ which contradicts our assumption that $j_1 n + s \ne j_2 n + t$. Consequently, $\alpha^e \mathbf{w}_{i_1 n + k_1}$ and $\alpha^f \mathbf{w}_{i_2 n + k_2}$ can have at most one position where they have identical nonzero components. It follows from the above proven facts that $\alpha^e \mathbf{w}_{i_1 n + k_1}$ and $\alpha^f \mathbf{w}_{i_2 n + k_2}$ differ in at least $cn - 1 = q - 2$ places. Hence, $\mathbf{W}$ satisfies the RD-constraint. ∎

Consider the special case for which $c = 1$ and $n = q - 1$. In this case, $\beta = \alpha$, $\delta = 1$ and

$$\mathbf{W} = \begin{bmatrix} 1 - 1 & 1 - \alpha & \cdots & 1 - \alpha^{q-2} \\ \alpha - 1 & \alpha - \alpha & \cdots & \alpha - \alpha^{q-2} \\ \vdots & \vdots & \ddots & \vdots \\ \alpha^{q-2} - 1 & \alpha^{q-2} - \alpha & \cdots & \alpha^{q-2} - \alpha^{q-2} \end{bmatrix}. \tag{6}$$

From (6) we see that every row (column) of $\mathbf{W}$ is the right (downward) cyclic-shift of the row (column) above it (on its left) multiplied by $\alpha$ and the first row is the right (downward) cyclic-shift of the last row (column) multiplied by $\alpha$. All the $q - 1$ entries in a row (or a column) of $\mathbf{W}$ are distinct elements in GF($q$). Each row (column) contains a 0-element. Therefore, in each row (or column), there is a nonzero element in GF($q$) that is not included.

For the special case with $c = q - 1$ and $n = 1$, we have the following RD-constrained matrix over GF($q$):

$$\mathbf{W} = \begin{bmatrix} \alpha^0 - 1 & \alpha - 1 & \cdots & \alpha^{q-2} - 1 \\ \alpha^{q-2} - 1 & \alpha^0 - 1 & \cdots & \alpha^{q-3} - 1 \\ \vdots & \vdots & \ddots & \vdots \\ \alpha - 1 & \alpha^2 - 1 & \cdots & \alpha^0 - 1 \end{bmatrix} \tag{7}$$

From (7), we see that every row of $\mathbf{W}$ is the right cyclic-shift of the row above and the first row is the right cyclic-shift of the last row. This matrix is exactly the same as the RD-constrained matrix given by Eq. (4) of [7] (with rows permuted) which was used as the base matrix for the major construction of QC-LDPC codes in [7]. Therefore, the construction of RD-constrained matrices presented in this paper is an expansion of the construction of the RD-constrained matrices (method 1) given in [7].

If we take the first columns from the submatrices, $\mathbf{W}_{0,0}, \mathbf{W}_{0,1}, \ldots, \mathbf{W}_{0,c-1}$, of $\mathbf{W}$ given by (4), then we obtain the following $n \times c$ submatrix of $\mathbf{W}$:

$$\mathbf{W}^* = \begin{bmatrix} 0 & \delta-1 & \cdots & \delta^{c-2}-1 \\ \beta-1 & \delta\beta-1 & \cdots & \delta^{c-1}\beta-1 \\ \vdots & \vdots & \ddots & \vdots \\ \beta^{n-1}-1 & \delta\beta^{n-1}-1 & \cdots & \delta^{c-1}\beta^{n-1}-1 \end{bmatrix} \tag{8}$$

$\mathbf{W}^*$ is exactly in the same form as that of the RD-constrained matrix $\mathbf{W}^{(3)}$ given by Eq. (6) in [8], except for the notations and that there is an extra column, $[-1, -1, \ldots, -1]^T$, in $\mathbf{W}^{(3)}$. Therefore, the RD-constrained matrix $\mathbf{W}$ given by (4) is an expansion of the RD-constrained matrix $\mathbf{W}^{(3)}$ given by Eq. (6) in [8].

## B. A Class of QC-LDPC Codes on Finite Fields

By array dispersion of $\mathbf{W}$ given by (4), we obtain the following $c \times c$ array of $n \times n$ subarrays of $(q-1) \times (q-1)$ CPMs and zero matrices over GF(2):

$$\mathbf{H} = [\mathbf{H}_{i,j}]_{0 \leq i < c, 0 \leq j < c}. \tag{9}$$

For $0 \leq i, j < c$,

$$\mathbf{H}_{i,j} = \begin{bmatrix} \mathbf{B}_{0,0}^{(i,j)} & \mathbf{B}_{0,1}^{(i,j)} & \cdots & \mathbf{B}_{0,n-1}^{(i,j)} \\ \mathbf{B}_{1,0}^{(i,j)} & \mathbf{B}_{1,1}^{(i,j)} & \cdots & \mathbf{B}_{1,n-1}^{(i,j)} \\ \vdots & \vdots & \ddots & \vdots \\ \mathbf{B}_{n-1,0}^{(i,j)} & \mathbf{B}_{n-1,1}^{(i,j)} & \cdots & \mathbf{B}_{n-1,n-1}^{(i,j)} \end{bmatrix} \tag{10}$$

is the array dispersion of the RD-constrained matrix $\mathbf{W}_{i,j}$, where $\mathbf{B}_{k,l}^{(i,j)} = \mathbf{B}(\delta^{j-i}\beta^k - \beta^l)$ is the matrix dispersion of the entry $\delta^{j-i}\beta^k - \beta^l$ at the $k$th row and $l$th column of $\mathbf{W}_{i,j}$. $\mathbf{B}_{k,l}^{(i,j)}$ is a $(q-1) \times (q-1)$ CPM if $\delta^{j-i}\beta^k - \beta^l \neq 0$ and a $(q-1) \times (q-1)$ ZM if $\delta^{j-i}\beta^k - \beta^l = 0$. From (9) and (10), we see that $\mathbf{H}$ is a $(q-1) \times (q-1)$ array of $(q-1) \times (q-1)$ CPMs and ZMs. Each row (or column) block of $\mathbf{H}$ consists of $q-2$ CPMs and one ZM. Therefore, $\mathbf{H}$ contains $q-1$ ZMs which lie on the main diagonal of $\mathbf{H}$. $\mathbf{H}$ is a $(q-1)^2 \times (q-1)^2$ matrix over GF(2) with both column and row weights $q-2$. Since $\mathbf{W}$ satisfies the RD-constraint, $\mathbf{H}$ satisfies the RC-constraint and can be used to construct RC-constrained QC-LDPC codes.

For any pair $(\gamma, \rho)$ of integers $\gamma$ and $\rho$ with $1 \leq \gamma, \rho \leq q$, let $\mathbf{H}(\gamma, \rho)$ be a $\gamma \times \rho$ subarray of $\mathbf{H}$, as a $(q-1) \times (q-1)$ array of CPMs and ZMs. $\mathbf{H}(\gamma, \rho)$ is a $\gamma(q-1) \times \rho(q-1)$ matrix over GF(2) which also satisfies the RC-constraint. The null space of $\mathbf{H}(\gamma, \rho)$ gives a binary

 

QC-LDPC code $\mathcal{C}_{qc}$ of length $\rho(q-1)$ with rate at least $(\rho-\gamma)/\rho$, whose Tanner graph has a girth of at least 6. For a given finite field GF($q$), the above construction gives a family of structurally compatible binary QC-LDPC codes.

If $\mathbf{H}(\gamma,\rho)$ does not contain any ZM of $\mathbf{H}$, $\mathbf{H}(\gamma,\rho)$, as a $\gamma(q-1) \times \rho(q-1)$ matrix over GF(2), has constant column weight $\gamma$ and constant row weight $\rho$. Then $\mathcal{C}_{qc}$ is a $(\gamma,\rho)$-regular QC-LDPC code with minimum distance at least $\gamma+1$. Note that the sum of the $q-1$ rows of a CPM gives an all-one $(q-1)$-tuple over GF(2). If we add all the $q-1$ rows of a row block of CPMs of $\mathbf{H}(\gamma,\rho)$, we obtain an all-one row vector $\mathbf{u} = (1\,1\ldots 1)$ of length $\rho(q-1)$ which is a codeword in the code $\mathcal{C}_{qc}^{\perp}$ spanned by the rows of $\mathbf{H}(\gamma,\rho)$ which is the dual code of $\mathcal{C}_{qc}$. Then the inner product each codeword of $\mathcal{C}_{qc}$ and the all-one vector $\mathbf{u}$ must be zero. This implies that every codeword in $\mathcal{C}_{qc}$ has even weight and hence the minimum weight of $\mathcal{C}_{qc}$ must be even. For even $\gamma$, $\gamma+1$ is odd. Then the minimum distance of $\mathcal{C}_{qc}$ must be at least $\gamma+2$. For odd $\gamma$, $\gamma+1$ is even. In this case, $\gamma+1$ gives a lower bound on the minimum distance of $\mathcal{C}_{qc}$. If $\mathbf{H}(\gamma,\rho)$ contains ZMs in some of its columns but not in all its columns, then $\mathbf{H}(\gamma,\rho)$, as a $\gamma(q-1) \times \rho(q-1)$ matrix, has two different column weights, $\gamma-1$ and $\gamma$. In this case, the RC-constraint ensures the minimum distance of the QC-LDPC code $\mathcal{C}_{qc}$ given by the null space of $\mathbf{H}(\gamma,\rho)$ is at least $\gamma$.

In the following, we use two examples to illustrate the above construction of QC-LDPC codes. For each code constructed, we compute its error performance over the AWGN channel with BPSK signaling decoded using the SPA [3],[6],[13] with no more than 50 iterations.

*Example 1:* Let GF($2^4$) be the field for code construction. Suppose we factor $2^4-1=15$ as the product of 3 and 5. Set $c=3$ and $n=5$. Let $\alpha$ be a primitive element of GF($2^4$). Set $\beta = \alpha^3$ and $\delta = \alpha^5$. Then the orders of $\beta$ and $\delta$ are 5 and 3, respectively. Form two cyclic subgroups of the multiplicative group of GF($2^4$) as follows: $\mathscr{G}_1 = \{\beta^0 = 1, \beta, \beta^2, \beta^3, \beta^4\}$ and $\mathscr{G}_2 = \{\delta^0 = 1, \delta, \delta^2\}$. Based on (3) and (4), we construct a $3 \times 3$ array $\mathbf{W}$ of $5 \times 5$ submatrices over GF($2^4$). $\mathbf{W}$ is a $15 \times 15$ RD-constrained matrix over GF($2^4$). Dispersing each nonzero entry of $\mathbf{W}$ into a binary $15 \times 15$ CPM and each zero entry into a $15 \times 15$ ZM, we obtain a $15 \times 15$ array $\mathbf{H}$ of CPMs and ZMs of size $15 \times 15$. For any pair of positive integers, $(\gamma,\rho)$, with $1 \le \gamma, \rho \le 15$, the null space of a $\gamma \times \rho$ subarray $\mathbf{H}(\gamma,\rho)$ of $\mathbf{H}$ gives a binary QC-LDPC code of length $15\rho$. Suppose we choose $\gamma = \rho = 15$. In this case, we use the entire array $\mathbf{H}$ as the parity-check matrix. It is a $225 \times 225$ matrix over GF(2) with both column and row weights

14. The null space of $\mathbf{H}$ gives a $(225, 147)$ QC-LDPC code with rate $0.653$. Since the column weight is $14$, the minimum distance is at least $15$. The error performances of this code over the binary-input AWGN channel decoded using the SPA with 5, 10 and 50 iterations are shown in Figure 1. We see that the decoding of this code converges very fast. At the block error rate (BLER) of $10^{-6}$, the code performs $0.9$ dB from the sphere packing bound. $\triangle\triangle$

*Example 2:* Let GF(379) be the code construction field. Suppose we factor $379 - 1 = 378$ as the product 6 and 63. Set $c = 6$ and $n = 63$. Let $\alpha$ be a primitive element of GF(379). Set $\beta = \alpha^6$ and $\delta = \alpha^{63}$. Then the orders of $\beta$ and $\delta$ are 63 and 6, respectively. Form two cyclic subgroups of GF(379): $\mathscr{G}_1 = \{\alpha^0 = 1, \alpha, \ldots, \alpha^{62}\}$ and $\mathscr{G}_2 = \{\delta^0 = 1, \delta, \ldots, \delta^5\}$. Based on these two subgroups, (3), (4), (9) and (10), we can construct a $378 \times 378$ array $\mathbf{H}$ of CPMs and ZMs of size $378 \times 378$. Take a $4 \times 32$ subarray $\mathbf{H}(4, 32)$ from $\mathbf{H}$, avoiding ZMs. $\mathbf{H}(4, 32)$ is a $1512 \times 12096$ matrix over GF(2) with column and row weights 4 and 32, respectively. The null space of this matrix gives a binary $(4, 32)$-regular $(12096, 10587)$ QC-LDPC code with rate $0.8752$. The error performance of this code over the binary-input AWGN channel decoded using the SPA with 10 and 50 iterations are shown in Figure 2. At the BER of $10^{-8}$, the code performs only 1 dB from the Shannon limit. We also see that decoding of this code converges fast. At a BER of $10^{-8}$, the gap between 10 and 50 iterations in performance is only $0.2$ dB. $\triangle\triangle$

### C. Masking

A set of binary CPMs in a chosen $\gamma \times \rho$ subarray $\mathbf{H}(\gamma, \rho) = [\mathbf{B}_{k,l}]$ of the array $\mathbf{H}$ given by (9) can be replaced by zero matrices. This replacement is referred to as *masking* [6], [7], [14], [15]. Masking results in a sparser matrix whose associated Tanner graph has fewer edges and hence fewer short cycles and probably a larger girth than that of the associated Tanner graph of the original $\gamma \times \rho$ subarray $\mathbf{H}(\gamma, \rho)$. To carry out masking, we first design a low density $\gamma \times \rho$ matrix $\mathbf{Z}(\gamma, \rho) = [z_{k,l}]$ over GF(2). Then we take the following matrix product: $\mathbf{M}(\gamma, \rho) = \mathbf{Z}(\gamma, \rho) \bigotimes \mathbf{H}(\gamma, \rho) = [z_{k,l}\mathbf{B}_{k,l}]$, where $z_{k,l}\mathbf{B}_{k,l} = \mathbf{B}_{k,l}$ for $z_{k,l} = 1$ and $z_{k,l}\mathbf{B}_{k,l} = \mathbf{0}$ (a $(q-1) \times (q-1)$ zero matrix) for $z_{k,l} = 0$. We call $\mathbf{Z}(\gamma, \rho)$ the masking matrix, $\mathbf{H}(\gamma, \rho)$ the base array and $\mathbf{M}(\gamma, \rho)$ the masked array, respectively. Since the base array $\mathbf{H}(\gamma, \rho)$ satisfies the RC-constraint, the masked array $\mathbf{M}(\gamma, \rho)$ also satisfies the RC-constraint, regardless of the masking matrix. Hence, the associated Tanner graph of the masked matrix $\mathbf{M}(\gamma, \rho)$ has a girth at least 6. The null space of the masked array $\mathbf{M}(\gamma, \rho)$ gives a new binary QC-LDPC code [6], [7], [15]. Masking can be either regular or irregular. Masking subarrays of $\mathbf{H}$ produces many more QC-LDPC codes.

*Example 3:* In this example, we construct a long irregular code and show how close it performs to the Shannon limit over the binary-input AWGN channel with iterative decoding. Let GF($2^9$) be the field for code construction. Suppose we factor $512 - 1 = 511$ as the product of $7 \times 73$. Set $c = 7$ and $n = 73$. Let $\alpha$ be a primitive element of GF($2^9$). Set $\beta = \alpha^7$ and $\delta = \alpha^{73}$. Form two cyclic subgroups of the multiplicative group of GF($2^9$), $\mathscr{G}_1 = \{\beta^0, \beta, \ldots, \beta^{72}\}$ and $\mathscr{G}_2 = \{\delta^0, \delta, \ldots, \delta^6\}$. Based on these two groups, (3), (4), (9) and (10), we construct an RC-constrained $511 \times 511$ array $\mathbf{H}$ of CPMs and ZMs of size $511 \times 511$ with the ZMs lying on the main diagonal of the array. Choose $\gamma = 63$ and $\rho = 126$. Take a $63 \times 126$ subarray $\mathbf{H}(63, 126)$ from the array $\mathbf{H}$, avoiding zero matrices. We will use this subarray as a base array for masking to construct an irregular code of rate $0.5$.

Consider the following degree distributions of variable nodes and check nodes of a Tanner graph optimally designed for an irregular code with rate $1/2$ and infinite length: $\lambda(X) = 0.4410X + 0.3603X^2 + 0.00171X^5 + 0.03543X^6 + 0.09331X^7 + 0.0204X^8 + 0.0048X^9 + 0.04305X^{29}$, and $\rho(X) = 0.00842X^7 + 0.99023X^8 + 0.00135X^9$, where the coefficient of $X^i$ represents the percentage of nodes with degree $i + 1$. Next, we construct a $63 \times 126$ matrix $\mathbf{Z}(63, 126)$ matrix over GF(2) with column and row weight distributions based on the above degree distributions. By computer search, we construct such a matrix with column and row weight distributions given in Table 1. Masking the $63 \times 126$ subarray $\mathbf{H}(63, 126)$ with $\mathbf{Z}(63, 126)$, we obtain a $63 \times 126$ masked array $\mathbf{M}(63, 126) = \mathbf{Z}(63, 126) \bigotimes \mathbf{H}(63, 126)$ of $511 \times 511$ CPMs and ZMs of size $511 \times 511$. It is a $32193 \times 64386$ matrix over GF(2) with column and row weight distributions close to the optimal degree distributions of the variable and check nodes of the Tanner graph for an irregular LDPC code of rate $0.5$ given above. The null space of $\mathbf{M}(63, 126)$ gives an irregular binary $(64386, 32193)$ QC-LDPC code. The error performance of this code with 50 iterations of the SPA is shown in Figure 3. We see that at the BER of $10^{-6}$, the code performs $0.55$ dB from the Shannon limit. Also included in Figure 3 is the performance of a $(64386, 32193)$ pseudo-random irregular QC-LDPC code constructed with the PEG-algorithm [16] based on the same node degree distributions, $\lambda(X)$ and $\rho(X)$, given above. We see that the algebraic code slightly outperforms its corresponding pseudo-random code. $\triangle\triangle$

## IV. Rank Analysis

In this section, we analyze the ranks of the parity-check matrices of a subclass of QC-LDPC codes constructed in Section III with $q = 2^m$, i.e., codes constructed based on GF($2^m$).

*Definition 1:* Let $\mathbf{A} = [a_{i,j}]$ and $\mathbf{B} = [b_{i,j}]$ be two $k \times n$ matrices over GF($q$). The *Hadamard product* of $\mathbf{A}$ and $\mathbf{B}$ is defined as their element-wise product $\mathbf{A} \circ \mathbf{B} = [a_{i,j} b_{i,j}]$ [17].

It is clear from the definition that Hadamard product $\mathbf{A} \circ \mathbf{B}$ of $\mathbf{A}$ and $\mathbf{B}$ is also a $k \times n$ matrix over GF($q$). If $\mathbf{B} = \mathbf{A}$, then $\mathbf{A}^{\circ 2} = \mathbf{A} \circ \mathbf{A} = [a_{i,j}^2]$. For any positive integer $l$, let $\mathbf{A}^{\circ l}$ denote the Hadamard product of $\mathbf{A}$ with itself $l$ times, i.e., $\mathbf{A}^{\circ l} = \mathbf{A} \circ \mathbf{A} \circ \ldots \circ \mathbf{A}$. Then $\mathbf{A}^{\circ l} = [a_{i,j}^l]$. We call $\mathbf{A}^{\circ l}$ the $l$th-fold Hadamard product of $\mathbf{A}$. For $l = 1$, $\mathbf{A}^{\circ 1} = \mathbf{A}$ and $\mathbf{A}^{\circ q} = \mathbf{A}$.

Let $\mathbf{G}$ be a matrix over GF($2^m$) and $\mathbf{M}$ be the binary $(2^m - 1)$-fold array dispersion of $\mathbf{G}$. Then $\mathbf{M}$ is an array of CPMs and/or ZMs over GF(2) of size $(2^m - 1) \times (2^m - 1)$. It has been proved in [18] that the rank of $\mathbf{M}$, denoted by $rank(\mathbf{M})$, can be expressed in terms of the ranks of the Hadamard products of $\mathbf{G}$, $\mathbf{G}^{\circ 1}, \mathbf{G}^{\circ 2}, \ldots, \mathbf{G}^{\circ(2^m-1)}$ as given in Theorem 3.

*Theorem 3:* Let $\mathbf{G}$ be a $k \times n$ matrix over GF($2^m$) and $\mathbf{M}$ be the binary $(2^m - 1)$-fold array dispersion of $\mathbf{G}$. Then the rank of the $k \times n$ array $\mathbf{M}$ of CPMs and/or ZMs of size $(2^m - 1) \times (2^m - 1)$ over GF(2) is equal to

$$rank(\mathbf{M}) = \sum_{l=1}^{2^m-1} rank(\mathbf{G}^{\circ l}). \tag{11}$$

For the simplicity of analysis, we consider the RD-constrained matrix $\mathbf{W}$ over GF($2^m$) given by (6). Since the characteristic of GF($2^m$) is 2, the subtraction "–" in (6) can be replaced by modulo-2 addition "+". Let $\mathscr{A} = \{0, 1, \ldots, 2^m - 2\}$ be an index set of order $2^m - 1$. Label the rows and columns of $\mathbf{W}$ of (6) in the order of $0, 1, \ldots, 2^m - 2$. Then,

$$\mathbf{W} = \left[\alpha^i + \alpha^j\right]_{i \in \mathscr{A}, j \in \mathscr{A}}. \tag{12}$$

Then, for any positive integer $l$, the $l$th-fold Hadamard product $\mathbf{W}^{\circ l}$ of $\mathbf{W}$ is given by

$$\mathbf{W}^{\circ l} = \left[(\alpha^i + \alpha^j)^l\right]_{i \in \mathscr{A}, j \in \mathscr{A}}. \tag{13}$$

Let $\mathbf{H}$ be the array dispersion of the RD-constrained matrix $\mathbf{W}$ given in the form of (12). It is $(2^m - 1) \times (2^m - 1)$ array of CPMs and ZMs of size $(2^m - 1) \times (2^m - 1)$. Corresponding to the column and row labeling of $\mathbf{W}$, we label the row and column blocks (CPMs and/or ZMs) of $\mathbf{H}$ in the order of $0, 1, \ldots, 2^m - 2$. For $1 \le \gamma \le 2^m - 1$ and $\rho = 2^m - 1$, let $\mathbf{H}(\gamma, 2^m - 1)$ be a $\gamma \times (2^m - 1)$ subarray of $\mathbf{H}$ that consists of $\gamma$ row blocks of $\mathbf{H}$. Without loss of generality, we take the first $\gamma$ row blocks of $\mathbf{H}$ to form $\mathbf{H}(\gamma, 2^m - 1)$ for the simplicity of notations and expressions. Let $\mathbf{W}(\gamma, 2^m - 1)$ be the first $\gamma$ rows of $\mathbf{W}$. Then $\mathbf{H}(\gamma, 2^m - 1)$ is the array dispersion of $\mathbf{W}(\gamma, 2^m - 1)$. It follows from Theorem 3 that the rank $rank(\mathbf{H}(\gamma, 2^m - 1))$ is given as follows:

$$rank(\mathbf{H}(\gamma, 2^m - 1)) = \sum_{l=1}^{2^m - 1} rank(\mathbf{W}^{\circ l}(\gamma, 2^m - 1)). \tag{14}$$

*Theorem 4:* For $1 \le l < 2^m$, let $\lambda_l$ be the number of odd integers in the $l$th row of the *Pascal's triangle* [19]. Then, for $1 \le \gamma \le 2^m - 1$, the rank of $\mathbf{W}^{\circ l}(\gamma, 2^m - 1)$ is given as follows:

$$rank\left(\mathbf{W}^{\circ l}\left(\gamma, 2^m - 1\right)\right) = \begin{cases} \min\left(\gamma, \lambda_l\right), & \text{for } 1 \le l < 2^m - 1, \\ \min\left(\gamma, \lambda_l - 2\right) = \min\left(\gamma, 2^m - 2\right), & \text{for } l = 2^m - 1. \end{cases} \tag{15}$$

*Proof:* Let $\mathscr{A}_\gamma$ be the subset of index set $\mathscr{A}$ which consists of the first $\gamma$ indices of $\mathscr{A}$. Then, the $\gamma \times (2^m - 1)$ submatrix $\mathbf{W}(\gamma, 2^m - 1)$ of $\mathbf{W}$ can be expressed as follows:

$$\mathbf{W}(\gamma, 2^m - 1) = \left[\alpha^i + \alpha^j\right]_{i \in \mathscr{A}_\gamma, j \in \mathscr{A}}.$$

For $1 \le l < 2^m$, the $l$th-fold Hadamard product $\mathbf{W}^{\circ l}(\gamma, 2^m - 1)$ of $\mathbf{W}(\gamma, 2^m - 1)$ is given by

$$\mathbf{W}^{\circ l}(\gamma, 2^m - 1) = \left[(\alpha^i + \alpha^j)^l\right]_{i \in \mathscr{A}_\gamma, j \in \mathscr{A}}.$$

Binomial expansion of $(\alpha^i + \alpha^j)^l$ results in the following expression:

$$\left(\alpha^i + \alpha^j\right)^l = \sum_{t=0}^{l} \binom{l}{t} \alpha^{i(l-t)} \alpha^{jt}. \tag{16}$$

Since the characteristic of GF($2^m$) is 2, $\binom{l}{t} = 1$ (modulo-2) if $\binom{l}{t}$ is odd and $\binom{l}{t} = 0$ (modulo-2) if $\binom{l}{t}$ is even. Let $t_1, t_2, \ldots, t_{\lambda_l}$ be the set of nonnegative integers for which the binomial coefficients $\binom{l}{t_1}, \binom{l}{t_2}, \ldots, \binom{l}{t_{\lambda_l}}$ are odd. Note that $\lambda_l$ is simply the number of odd integers in the Pascal's triangle at the $l$th level and $\lambda_l \le l + 1$. Also note that $\binom{l}{0} = \binom{l}{l} = 1$, thus $t_1 = 0$ and $t_{\lambda_l} = l$. Then the binomial expression of (16) is reduced to the following form:

$$\left(\alpha^i + \alpha^j\right)^l = \alpha^{il} + \alpha^{i(l-t_2)}\alpha^{jt_2} + \alpha^{i(l-t_3)}\alpha^{jt_3} + \ldots + \alpha^{i(l-t_{\lambda_l - 1})}\alpha^{jt_{\lambda_l - 1}} + \alpha^{jl}. \tag{17}$$

There are two cases to be considered. First, we consider the case for which $1 \le l < 2^m - 1$. Based on the expressions of (17) and $\mathbf{W}^{\circ l}(\gamma, 2^m - 1) = \left[(\alpha^i + \alpha^j)^l\right]_{i \in \mathscr{A}_\gamma, j \in \mathscr{A}}$, the $l$th-fold

Hadamard product $\mathbf{W}^{\circ l}(\gamma, 2^m - 1)$ of $\mathbf{W}(\gamma, 2^m - 1)$ can be put into the following form:

$$
\mathbf{W}^{\circ l}(\gamma, 2^m-1) = \underbrace{\begin{bmatrix} (\alpha^0)^l & (\alpha^0)^{l-t_2} & \cdots & 1 \\ (\alpha^1)^l & (\alpha^1)^{l-t_2} & \cdots & 1 \\ \vdots & \vdots & \vdots & \vdots \\ (\alpha^{\gamma-1})^l & (\alpha^{\gamma-1})^{l-t_2} & \cdots & 1 \end{bmatrix}}_{\mathbf{L}_{\gamma \times \lambda_l}} \underbrace{\begin{bmatrix} 1 & 1 & \cdots & 1 \\ (\alpha^0)^{t_2} & (\alpha^1)^{t_2} & \cdots & (\alpha^{2^m-2})^{t_2} \\ \vdots & \vdots & \vdots & \vdots \\ (\alpha^0)^l & (\alpha^1)^l & \cdots & (\alpha^{2^m-2})^l \end{bmatrix}}_{\mathbf{R}_{\lambda_l \times (2^m-1)}}
$$
(18)

$$
= \begin{bmatrix} \mathbf{L}_{\gamma \times \lambda_l} & \mathbf{0}_{\gamma \times (2^m-1-\lambda_l)} \end{bmatrix} \begin{bmatrix} \mathbf{R}_{\lambda_l \times (2^m-1)} \\ \tilde{\mathbf{R}}_{(2^m-1-\lambda_l) \times (2^m-1)} \end{bmatrix}.
$$

The matrix $\mathbf{R}_{\lambda_l \times (2^m-1)}$ consists of $\lambda_l$ rows of the transpose $\mathbf{V}^T$ of the following *Vandermonde matrix*:

$$
\mathbf{V} = \begin{bmatrix} (\alpha^0)^{2^m-2} & \cdots & (\alpha^0)^1 & 1 \\ (\alpha^1)^{2^m-2} & \cdots & (\alpha^1)^1 & 1 \\ \vdots & \cdots & \vdots & \vdots \\ (\alpha^{2^m-2})^{2^m-2} & \cdots & (\alpha^{2^m-2})^1 & 1 \end{bmatrix}_{(2^m-1) \times (2^m-1)}.
$$
(19)

The matrix $\tilde{\mathbf{R}}$ consists of all the other rows of $\mathbf{V}^T$. Therefore $\mathbf{V}^T = [\mathbf{R}^T \ \tilde{\mathbf{R}}^T]^T$. The matrix $\mathbf{0}_{\lambda_l \times (2^m-1-\lambda_l)}$ is a $\lambda_l \times (2^m-1-\lambda_l)$ ZM. Since $\mathbf{L}_{\gamma \times \lambda_l}$ is $\gamma \times \lambda_l$ submatrix of the Vandermonde matrix $\mathbf{V}$, $rank(\mathbf{L}_{\gamma \times \lambda_l}) = \min(\gamma, \lambda_l)$. Since the rank of $\mathbf{V}^T$ is $2^m-1$, then it follows from (18) that $rank(\mathbf{W}^{\circ l}(\gamma, 2^m-1)) = rank(\mathbf{L}_{\gamma \times \lambda_l}) = \min(\gamma, \lambda_l)$. This proves the first part of the theorem for the case $1 \le l < 2^m - 1$.

Now, we consider the case for $l = 2^m - 1$. From (18), we can see that when $l = 2^m - 1$, the first column of $\mathbf{L}_{\gamma \times \lambda_l}$ becomes an all-1 vector, which is the same as the last column of $\mathbf{L}_{\gamma \times \lambda_l}$. Also we note that the first row of $\mathbf{R}_{\lambda_l \times (2^m-1)}$ is an all-1 vector which is the same the last row of $\mathbf{R}_{\lambda_l \times (2^m-1)}$. For $l = 2^m - 1$ and $\lambda_l = 2^m$. we have,

$$
\mathbf{W}^{\circ (2^m-1)}(\gamma, 2^m-1) = \underbrace{\begin{bmatrix} 1 & (\alpha^0)^{2^m-2} & \cdots & \alpha^0 & 1 \\ 1 & (\alpha^1)^{2^m-2} & \cdots & \alpha^1 & 1 \\ \vdots & \vdots & \vdots & \vdots & \vdots \\ 1 & (\alpha^{\gamma-1})^{2^m-2} & \cdots & \alpha^{\gamma-1} & 1 \end{bmatrix}}_{\mathbf{L}_{\gamma \times 2^m}} \underbrace{\begin{bmatrix} 1 & 1 & \cdots & 1 \\ (\alpha^0)^1 & (\alpha^1)^1 & \cdots & (\alpha^{2^m-2})^1 \\ \vdots & \vdots & \vdots & \vdots \\ (\alpha^0)^{2^m-2} & (\alpha^1)^{2^m-2} & \cdots & (\alpha^{2^m-2})^{2^m-2} \\ 1 & 1 & \cdots & 1 \end{bmatrix}}_{\mathbf{R}_{2^m \times (2^m-1)}}
$$

$$
= \underbrace{\begin{bmatrix} (\alpha^0)^{2^m-2} & \cdots & \alpha^0 \\ (\alpha^1)^{2^m-2} & \cdots & \alpha^1 \\ \vdots & \vdots & \vdots \\ (\alpha^{\gamma-1})^{2^m-2} & \cdots & \alpha^{\gamma-1} \end{bmatrix}}_{\mathbf{L}_{\gamma \times (2^m-2)}} \underbrace{\begin{bmatrix} (\alpha^0)^1 & (\alpha^1)^1 & \cdots & (\alpha^{2^m-2})^1 \\ \vdots & \vdots & \vdots & \vdots \\ (\alpha^0)^{2^m-2} & (\alpha^1)^{2^m-2} & \cdots & (\alpha^{2^m-2})^{2^m-2} \end{bmatrix}}_{\mathbf{R}_{(2^m-2) \times (2^m-1)}}
$$
(20)

Since $\mathbf{R}_{(2^m-2)\times(2^m-2)}$ is a square submatrix of the Vandermonde matrix $\mathbf{V}$, $rank(\mathbf{R}_{(2^m-2)\times(2^m-2)}) = 2^m - 2$. Then, it follows from (20) that

$$rank(\mathbf{W}^{\circ(2^m-1)}(\gamma, 2^m - 1)) = \min(\gamma, 2^m - 2).$$

This completes the proof of the theorem. ∎

A combinational expression for the rank of a $\gamma \times (2^m-1)$ subarray $\mathbf{H}(\gamma, 2^m-1)$ of the array $\mathbf{H}$ given by (6) can be derived.

*Theorem 5:* For $q = 2^m$, $1 \leq \gamma \leq 2^m - 1$, let $t_\gamma$ be the largest positive integer such that $2^{t_\gamma} \leq \gamma < 2^{t_\gamma+1}$. Then

$$rank\left(\mathbf{H}\left(\gamma, 2^m - 1\right)\right) = \begin{cases} \gamma\left(2^m - 1\right) - \sum\limits_{t=1}^{t_\gamma} \binom{m}{t}\left(\gamma - 2^t\right), & \text{for } 1 \leq \gamma < 2^m - 1, \\ 3^m - 3, & \text{for } \gamma = 2^m - 1. \end{cases} \quad (21)$$

*Proof:* It follows from (15) that for $1 \leq \gamma < 2^m - 1$, $rank\left(\mathbf{W}^{\circ l}\left(\gamma, 2^m - 1\right)\right) = \min\left(\gamma, \lambda_l\right)$, where $1 \leq l \leq 2^m - 1$. Then $rank\left(\mathbf{H}\left(\gamma, 2^m - 1\right)\right) = \sum\limits_{l=1}^{2^m-1} \min\left(\gamma, \lambda_l\right)$.

First, we prove the combinatorial expression for the case $1 \leq \gamma < 2^m - 1$. Label the Pascal's triangle from level-0. For $0 \leq l < 2^m$, the $l$th level of the Pascal's triangle consists of the following binomial coefficients: $\binom{l}{0} = 1$, $\binom{l}{1}$, $\binom{l}{2}$, ..., $\binom{l}{l-1}$, $\binom{l}{l} = 1$. An integer $l$ with $0 \leq l < 2^m$ can be expressed in the following radix-2 form: $l = a_0 + a_1 2 + a_2 2^2 + \ldots + a_{m-1}2^{m-1}$, where $a_i = 0$ or $1$ for $0 \leq i < m$. The sum $w(l) = \sum\limits_{i=0}^{m-1} a_i$ is called the *radix-2 weight* of the integer $l$. It is clear that $0 \leq w(l) \leq m$. Then $\lambda_l = 2^{w(l)}$. We readily see that $\gamma < \lambda_l$ if $t_\gamma < w(l)$ and $\lambda_l \leq \gamma$ if $w(l) \leq t_\gamma$.

Let $\mathscr{B}_0 = \{1, 2, \ldots, 2^m - 1\}$. Then the sum $\sum_{l=1}^{2^m-1} \min(\gamma, \lambda_l)$ can be put into the following form:

$$\sum_{l=1}^{2^m-1} \min(\gamma, \lambda_l) = \sum_{l=1}^{2^m-1} \min(\gamma, 2^{w(l)}) = \sum_{l \in \mathscr{B}_0, t_\gamma < w(l)} \gamma + \sum_{l \in \mathscr{B}_0, w(l) \leq t_\gamma} 2^{w(l)}.$$

The number of integers in $\mathscr{B}_0$ that have radix-2 weight $t$ with $0 \leq t \leq m$ is $\binom{m}{t}$. Then the above equality can be put in the following combinatorial form:

$$\sum_{l=1}^{2^m-1} \min(\gamma, \lambda_l) = \sum_{t=t_\gamma+1}^{m} \binom{m}{t}\gamma + \sum_{t=1}^{t_\gamma} \binom{m}{t}2^t = \gamma\sum_{t=1}^{m}\binom{m}{t} - \sum_{t=1}^{t_\gamma}\binom{m}{t}\left(\gamma - 2^t\right)$$

$$= \gamma(2^m - 1) - \sum_{t=1}^{t_\gamma}\binom{m}{t}\left(\gamma - 2^t\right).$$

 

This gives the first part of (21).

For the case $\gamma = 2^m - 1$, it follows from (14), (15) and $\lambda_{2^m-1} = 2^m$ that

$$rank\,(\mathbf{H}) = rank\,(\mathbf{H}\,(2^m - 1, 2^m - 1)) = \left(\sum_{l=1}^{2^m-2} \lambda_l\right) + 2^m - 2 = \left(\sum_{l=1}^{2^m-1} \lambda_l\right) - 2 \quad (22)$$

It is known that the total number of odd integers in the Pascal's triangle of $2^m$ levels (labeled from 0 to $2^m - 1$) is $3^m$ [19]. Since at the 0th level of the Pascal's triangle, there is a single odd integer which is "1", therefore the rank of $\mathbf{H}(2^m - 1, 2^m - 1)$ is $3^m - 3$. This gives the second equality of (21). ∎

*Example 4:* Let GF($2^6$) be the field for code construction. Based on this field, we construct a $63 \times 63$ RD-constrained matrix $\mathbf{W}$ over GF($2^6$) in the form given by (6). Array dispersion of $\mathbf{W}$ results in a $63 \times 63$ array $\mathbf{H}$ of CPMs and ZMs of size $63 \times 63$. Choose $\gamma = 6$. Suppose we take the first 6 rows of $\mathbf{H}$ to form a $6 \times 63$ subarray $\mathbf{H}(6, 63)$. $\mathbf{H}(6, 63)$ is a $378 \times 3969$ matrix over GF(2) with constant row weight 32 and two different column weights, 5 and 6. To determine the rank of $\mathbf{H}(6, 63)$, we apply Theorem 5. First, we find that $t_6 = 2$. Using the first combinatorial expression given by (21), we find that $rank(\mathbf{H}(6, 63)) = 324$. Hence the null space of $\mathbf{H}(6, 63)$ gives a $(3969, 3645)$ near-regular QC-LDPC code with rate $0.9183$. The performance of this code with 50 iterations of the SPA is shown in Figure 4. At the BLER of $10^{-4}$, the code performs $0.75$ dB from the sphere packing bound. At the BER of $10^{-6}$, the code performs $1.2$ dB from the Shannon limit. For comparison, a corresponding near-regular pseudo-random $(3969, 3645)$ QC-LDPC code is constructed with the PEG-algorithm. Its error performance is also included in Figure 4. We see that the algebraic $(3969, 3645)$ code outperforms its corresponding pseudo-random code. △△

*Example 5:* We use GF($2^7$) for code construction. Based on this field, we construct an RD-constrained matrix $\mathbf{W}$ over GF($2^7$) in the form given by (6). Dispersing $\mathbf{W}$, we obtain a $127 \times 127$ array $\mathbf{H}$ of CPMs and ZMs of size $127 \times 127$. Choose $\gamma = 6$. Suppose we take the first 6 rows of $\mathbf{H}$ to form a $6 \times 127$ subarray $\mathbf{H}(6, 127)$. $\mathbf{H}(6, 127)$ is a $762 \times 16129$ matrix over GF(2). Based on Theorem 5 and the first expression (21), we find that $t_6 = 2$ and $rank(\mathbf{H}(6, 127)) = 692$. Hence the null space of $\mathbf{H}$ gives a $(16129, 15437)$ QC-LDPC code with rate $0.9571$. The performance of this code with 50 iterations of the SPA is shown in Figure 5. At the BERs of $10^{-6}$ and $10^{-8}$, the code performs $0.8$ dB and $0.95$ dB from the Shannon limit, respectively. For comparison, a corresponding near-regular pseudo-random $(16129, 15437)$ QC-LDPC code is constructed with the PEG-algorithm. Its error performance is also included in Figure 5. We see that the algebraic code slightly outperforms its corresponding pseudo-random code. $\triangle\triangle$

## V. A Special Subclass of RC-Constrained QC-LDPC Codes

An RC-constrained $(\gamma, \rho)$-regular LDPC code whose parity-check matrix has column weight $\gamma$ is one-step majority-logic decodable and is capable of correcting $\lfloor \gamma/2 \rfloor$ or fewer errors with one-step majority-logic decoding (OSMLGD) [5], [6]. OSMLGD is one of the simplest hard-decision decoding methods which requires only binary logical operations. For an RC-constrained $(\gamma, \rho)$-regular LDPC code to be effective with OSMLGD, its parity-check matrix must have a reasonably large column weight $\gamma$.

For a given field GF($q$), let $\mathcal{C}_{qc,f}$ be the QC-LDPC code generated by the null space of the full RC-constrained array $\mathbf{H}$ obtained by array dispersion of the RD-constrained base matrix $\mathbf{W}$ given by (4). The subscript "$f$" of $\mathcal{C}_{qc,f}$ stands for "full array". Since the column weight of $\mathbf{H}$ is $q-2$, the code $\mathcal{C}_{qc,f}$ is capable of correcting $\lfloor (q-2)/2 \rfloor$ or fewer errors with the OSMLGD. For $q = 2^m$, it follows from the second expression of (21) (Theorem 5) that the rank of the full array $\mathbf{H}$ is $3^m - 3$. In this case, $\mathcal{C}_{qc,f}$ is an RC-constrained QC-LDPC code with the following parameters: 1) length $(2^m-1)^2$; 2) number of parity-check symbols $3^m-3$; 3) minimum distance at least $2^m-1$: and 4) OSMLGD error-correction capability $2^{m-1}-1$. Since the number of rows of $\mathbf{H}$ is $(2^m-1)^2$ and the rank of $\mathbf{H}$ is $3^m-3$, $\mathbf{H}$ has $(2^m-1)^2-3^m+3$ redundant (or linearly dependent) rows. For $m \geq 3$, $\mathbf{H}$ has a large row redundancy.

The code $\mathcal{C}_{qc,f}$ given by the full array $\mathbf{H}$, not only performs well with iterative decoding using the SPA but also provides good error performance when decoded using the iterative binary message-passing decoding algorithm (IBMPDA) presented in [20] with significant reduction in decoding complexity. The IBMPDA presented in [20] requires only integer additions and binary logical operations. The number of integer additions required per iteration in decoding $\mathcal{C}_{qc,f}$ is equal to the number of 1-entries in $\mathbf{H}$ which is $(q-2)(q-1)^2$. It is shown in [20] that this IBMPDA outperforms all the known existing weighted bit flipping (WBF) decoding algorithms with much less computational complexity and performs close to the SPA.

*Example 6:* Consider the $63 \times 63$ RC-constrained array $\mathbf{H}$ of CPMs and ZMs of size $63 \times 63$ constructed based on GF($2^6$) given in Example 4. It is a $3969 \times 3969$ matrix over GF(2) with both column and row weights 62. Using the second expression of (21) given in Theorem 5, we find that the rank of $\mathbf{H}$ is 726. The null space of $\mathbf{H}$ gives a $(3969, 3243)$ RC-constrained QC-LDPC code $\mathcal{C}_{qc,f}$ with rate $0.8171$ and minimum distance at least $63$. The error performance of this code over the AWGN channel decoded using the SPA with 5, 10 and 50 iterations is shown in Figure 6. We see that the decoding of this code converges very fast. The performance gap between 10 and 50 iterations is negligible and the performance gap between 5 and 50 iterations is less than $0.2$ dB at the BER of $10^{-6}$. At the BLER of $10^{-5}$, the code performs $1.2$ dB from the sphere packing bound. Also included in Figure 6 are the performances of the code decoded with the IBMPDA presented in [20] and the OSMLGD. We see that at the BER of $10^{-6}$, the IBMPDA performs only $0.6$ dB from the SPA. With OSMLGD, the code is capable of correcting 31 or fewer errors. $\triangle\triangle$

## VI. Conclusion

In this paper, we first presented a large class of arrays of circulant permutation matrices that are constructed based on cyclic subgroups of finite fields. Based on this class of arrays of circulant permutation matrices, we constructed a large class of new QC-LDPC codes whose Tanner graphs have girth of at least $6$. Then, we analyzed the ranks of the parity-check matrices of codes constructed based on finite fields of characteristic $2$ and derived combinatorial expressions for these ranks. Experimental results show that the codes constructed perform well over the binary-input AWGN channel with iterative decoding using the SPA and they outperform the corresponding pseudo-random QC-LDPC codes constructed with the PEG-algorithm. In the paper, we also identified a subclass of constructed QC-LDPC codes that have large minimum distances. Decoding of codes in this subclass with the SPA converges very fast. Furthermore, we showed that, when decoded with the binary message-passing decoding algorithm recently devised in [20], codes in this subclass give close to the SPA performance with enormous reduction in decoding complexity. These codes may find applications in communication or storage systems where good error performance, fast decoding convergence, simple decoders and low error-floors are required. We also showed that the class of RD-constrained matrices constructed in this paper contains the first class of RD-constrained matrices given in [7] and the third class of RD-constrained matrices given in [8] as special subclasses.

The technique used to analyze the ranks of parity-check matrices of QC-LDPC code on cyclic subgroups of finite fields in this paper can be used to analyze the ranks of parity-check matrices of QC-LDPC codes on additive subgroups of finite fields presented in [9].

TABLE I

COLUMN AND ROW WEIGHT DISTRIBUTIONS OF THE MASKING MATRIX $\mathbf{Z}(63, 126)$ OF EXAMPLE 3

| Column Weight Distribution | | Row Weight Distribution | |
|---|---|---|---|
| Column weight | No. of columns | Row weight | No. of rows |
| 2 | 57 | 8 | 11 |
| 3 | 44 | 9 | 52 |
| 8 | 20 | | |
| 30 | 5 | | |

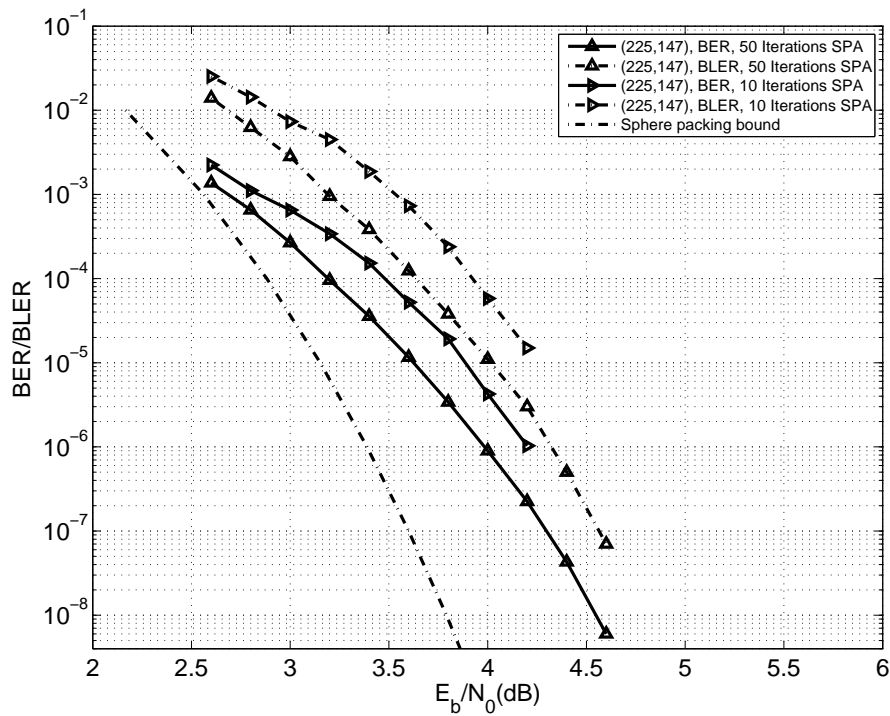

Fig. 1.   The error performance of the $(225, 147)$ QC-LDPC code given in Example 1 over the AWGN channel.

                                                                                                                

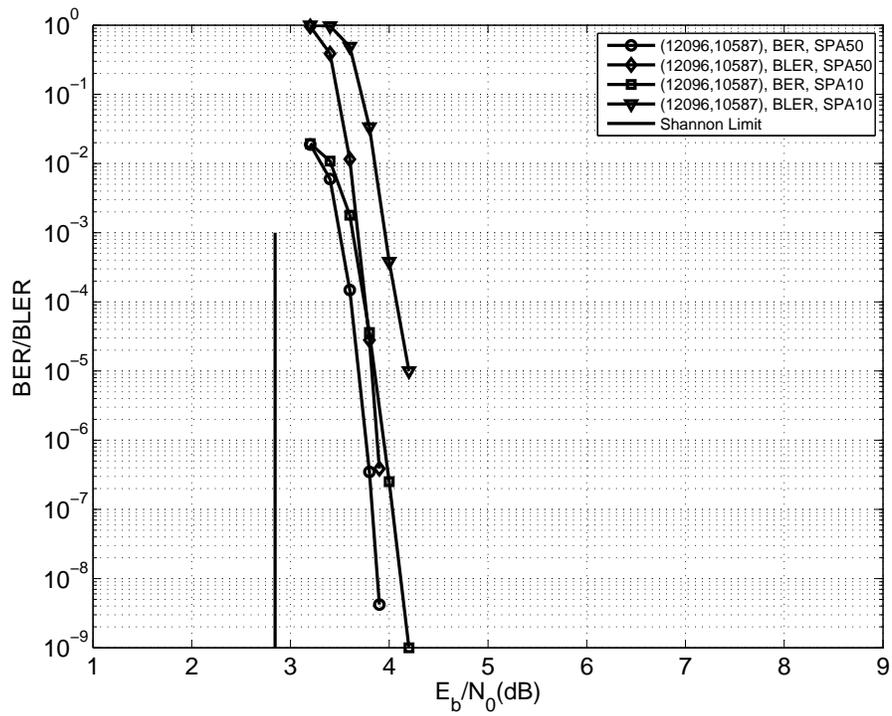

Fig. 2.   The error performance of the (12096, 10587) QC-LDPC code given in Example 2 over the AWGN channel.

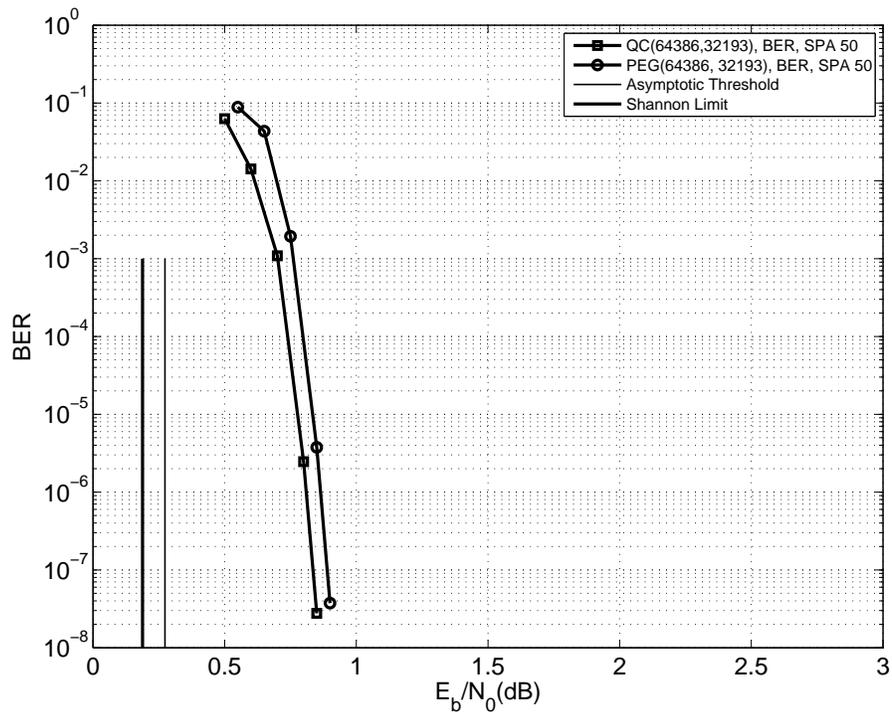

Fig. 3. The error performance of the (64386, 32193) QC-LDPC code given in Example 3 over the AWGN channel.

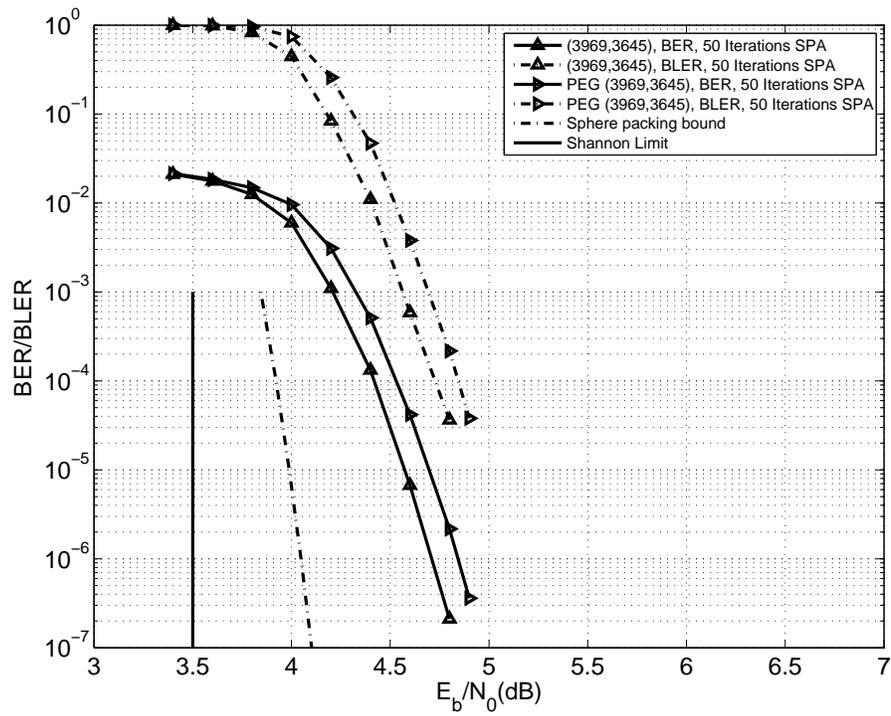

Fig. 4. The error performance of the (3969, 3645) QC-LDPC code given in Example 4 over the AWGN channel.

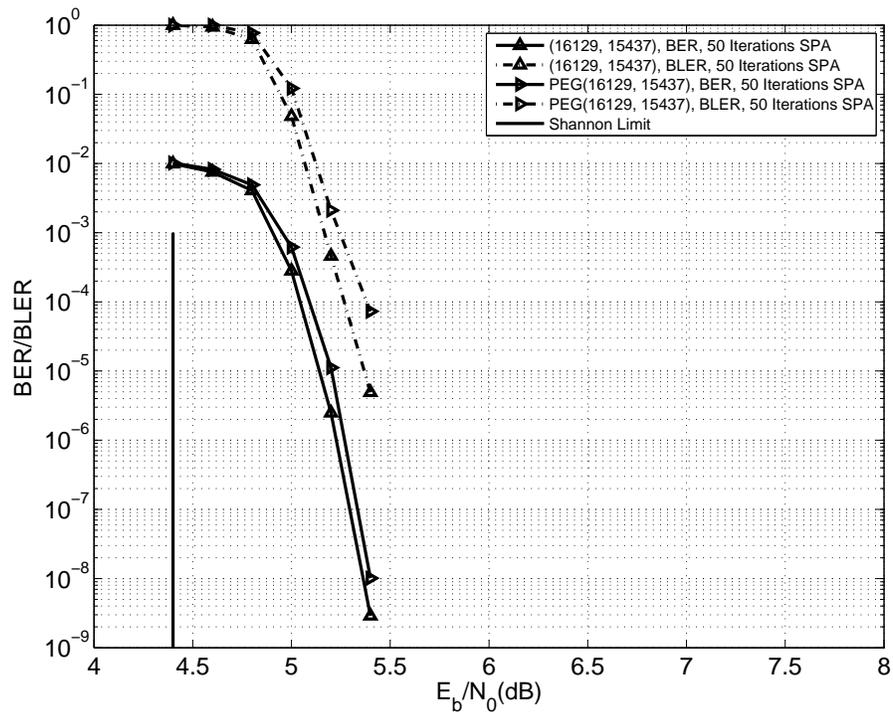

Fig. 5. The error performance of the (16129, 15437) QC-LDPC code given in Example 5 over the AWGN channel.

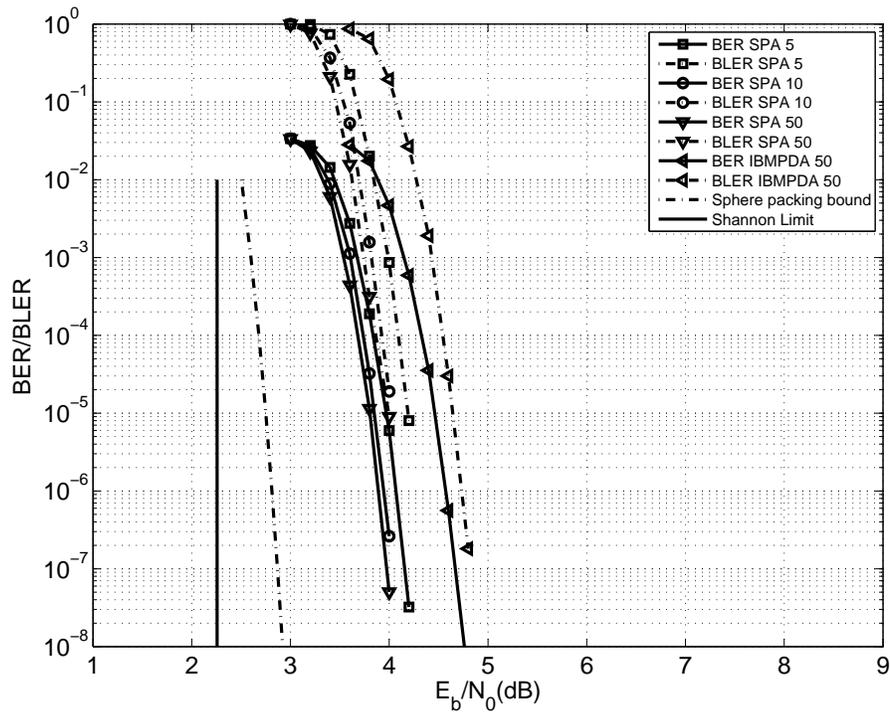

Fig. 6. The error performances of the $(3969, 3243)$ QC-LDPC code given in Example 6 over the AWGN channel decoded with the IBMPDA and the SPA.